\documentclass[useAMS,usenatbib,usegraphicx]{mn2e}
\usepackage{epsf}
\usepackage{amssymb,amsmath}

\title[Lensed quasars and supernovae in future surveys]
{Gravitationally lensed quasars and supernovae\\
in future wide-field optical imaging surveys}

\author[M.~Oguri and P.~J.~Marshall]
{Masamune~Oguri$^{1,2}$\thanks{E-mail: masamune.oguri@nao.ac.jp} 
and Philip~J.~Marshall$^{2,3}$\\
$^1$Division of Theoretical Astronomy, National Astronomical
Observatory of Japan, 2-21-1 Osawa, Mitaka, Tokyo 181-8588, Japan.\\ 
$^2$Kavli Institute for Particle Astrophysics and Cosmology, 
Stanford University, 2575 Sand Hill Road, Menlo Park, CA
94025, USA.\\
$^3$Physics Department, University of California, Santa Barbara, CA
93601, USA.
} 

\begin{document}

\date{\today}

\voffset- .5in

\pagerange{\pageref{firstpage}--\pageref{lastpage}} \pubyear{}

\maketitle

\label{firstpage}

\begin{abstract}
Cadenced optical imaging surveys in the next decade will be capable of
detecting time-varying galaxy-scale strong gravitational lenses in
large numbers, increasing the size of the statistically well-defined
samples of multiply-imaged quasars by two orders of magnitude, and
discovering the first strongly-lensed supernovae. We carry out a
detailed calculation of the likely yields of several planned surveys,
using realistic distributions for the lens and source properties and
taking magnification bias and image configuration detectability into
account. We find that upcoming wide-field synoptic surveys should detect
several thousand lensed quasars. In particular, the Large Synoptic
Survey Telescope (LSST) should find more than some 8000 lensed
quasars, some 3000 of which will have well-measured time delays. The LSST
should also find some 130 lensed supernovae during the 10-year survey
duration, which is compared with $\sim 15$ lensed supernovae predicted
to be found by a deep, space-based supernova survey done by the Joint
Dark Energy Mission (JDEM). We compute the quad fraction in each
survey, predicting it to be $\sim 15\%$ for the lensed quasars and
$\sim 30\%$ for the lensed supernovae. 
Generating a mock catalogue of around 1500 well-observed double-image
lenses, as could be derived from the LSST survey, we compute the
available precision on the Hubble constant and the dark energy
equation parameters for the time delay distance experiment (assuming
priors from Planck): the predicted marginalised 68\% confidence
intervals are $\sigma(w_0)=0.15$, $\sigma(w_{\rm a})=0.41$, and
$\sigma(h)=0.017$. While this is encouraging in the sense that these
uncertainties are only 50\% larger than those predicted for a
space-based type-Ia supernova sample, we show how the dark energy
figure of merit degrades with decreasing knowledge of the the lens
mass distribution.  
\end{abstract}

\begin{keywords}
cosmological parameters 
--- cosmology: theory 
--- gravitational lensing
\end{keywords}

\section{Introduction}
\label{sec:intro}

The discovery of strong gravitational lensing in Q0957$+$561 
\citep{walsh79} opened up the possibility of using strong lens systems
to study cosmology and astrophysics. Roughly a few hundred strong
lens systems produced by massive galaxies have been discovered to
date, with quasars \citep[e.g.,][]{inada08}, radio sources
\citep[e.g.,][]{myers03,browne03}, and galaxies
\citep[e.g.,][]{cabanac07,faure08,bolton08} appearing as source
populations. These samples of strong lenses have been extensively used
to constrain dark energy, the density profile of lensing galaxies, and
the evolution of massive ellipticals \citep[see, e.g.,][for a
review]{kochanek06a}.  

It is expected that large new samples of strong lenses will be
obtained in future wide-field imaging surveys
\citep[e.g.,][]{kuhlen04,marshall05}. For instance, the Large Synoptic
Survey Telescope \citep[LSST;][]{ivezic08} will observe a
20,000~deg$^2$ region with a final limiting magnitude of $r\sim 27.5$, 
which is considerably wider and deeper than existing optical imaging
surveys such as the Sloan Digital Sky Survey \citep[SDSS;][]{york00}. 
More importantly, future imaging surveys such as LSST's will pay particular
attention to the acquisition of time-domain data, mostly in order to
study transient objects including supernovae (SNe). 

In this paper, we present detailed predictions of the numbers of 
time-variable sources we can expect to be
strongly lensed in ongoing and future time
domain optical imaging surveys. We focus on time-variable sources for
two reasons. First,  time domain data enable us to identify strong
lenses by taking advantage of time variability \citep{pindor05}. 
\citet{kochanek06b} proposed to find strong lenses by looking for
``extended variable sources'' in time-domain data, and argued that
this technique should be very efficient due to the low levels of
contamination in the difference images. Second, time-variable sources 
allow us to measure time delays between multiple images; these time delays
contain
rich information on the lens potentials as well as cosmology
\citep[e.g.,][]{witt00,kochanek02,saha03,schechter05,oguri07a,
congdon08,congdon10,keeton09,keeton10}. 

The obvious and traditional 
example of such a strongly lensed time-variable source is a
quasar (QSO). Of $\sim 100$ gravitationally lensed quasars
currently known, time delays have been measured for only 
$\lesssim 20$ \citep[see][for a recent compilation]{oguri07a}.  
A major advance offered by the upcoming imaging surveys is their
potential not only to discover new lenses but also monitor them as
well. Some of the survey telescopes we will consider do have larger
static surveys planned, but we do not consider them on the grounds
that if they are wide enough to be competitive, they will be limited
by the required monitoring follow-up. Synoptic surveys (such as those
designed to discover supernovae) should provide some estimate of the
lens time delays -- we are careful to point out where this may not be
the case. 

The scope of several planned surveys allows us also to consider SNe as
time-variable sources. While a distant SN highly magnified by a
foreground massive cluster has recently been identified
\citep{stanishev09,goobar09}, no multiply-imaged SN has yet been
discovered. However, the possibility of discovering 
such strongly-lensed SNe by future time-domain surveys such as
the Joint Dark Energy Mission (JDEM) and the LSST has been pointed out
\citep{porciani00,holz01,goobar02,oguri03a,mortsell05}. A particularly
enticing feature of lensed SNe is that, if they are type-Ia, the
``standardizable candle'' nature of the peak brightness allows a
direct determination of the lensing magnification factor, which breaks
the degeneracy between the lens potential and the Hubble constant
\citep{oguri03b}.  

Indeed, we expect any large sample of time delay lenses to be useful
in constraining cosmological parameters. The idea of using strong
gravitational lens systems with time-variable sources to measure the
Hubble constant is an old one \citep{refsdal64}; in fact, the time
delays between images provide a way of measuring the ``time-delay
distance'' to the lens. This quantity is a combination of the angular
diameter distances to the lens, source, and between the two, and while
primarily sensitive to $H_0$, it does also depend on the other 
cosmological parameters 
\citep[see e.g.,][for a recent high-precision measurement]{suyu10}. 
Attempts to measure the Hubble constant from the statistical analysis
of the current lensed quasar sample \citep{saha06,oguri07a,cole08}
have yielded values consistent with constraints from other
cosmological probes
\citep[e.g.,][]{freedman01,bonamente06,komatsu09,riess09}. In future,
larger samples covering a wide range of lens and source redshift
permutations should provide interesting constraints on the
dark energy equation of state as well as Hubble constant
\citep{lewis02,linder04,dobke09,coe09}. 
In this work, we revisit the forecast constraints on dark energy parameters
for the LSST, using new, realistic predicted 
distributions of strong lenses.  

This paper is organised as follows. In Section~\ref{sec:calc} we
describe the ingredients of our lens abundance calculation, including
our assumptions about both the lens and source populations. We then
present our predictions for various survey depths and areas in
Section~\ref{sec:survey}. We then present Monte Carlo realisations of
some fiducial survey lens catalogues in Section~\ref{sec:mock}, and use
these to explore the potential of some of these in constraining
cosmological parameters via the time delay distances to the lenses in
Section~\ref{sec:cosmo}. After a brief discussion
(Section~\ref{sec:discuss}) we conclude in Section~\ref{sec:concl}. 
We take as the parameters of a fiducial cosmological model
$\Omega_{\rm m}=0.26$, $\Omega_{\rm DE}=0.74$, $h=0.72$, $w_0=-1$ and 
$w_{\rm a}=0$. We denote
the angular diameter distance from observer to lens as $D_{\rm l}$,
from observer to source as $D_{\rm s}$, and from lens to source as 
$D_{\rm ls}$.

\section{Predicting the numbers of lensed quasars and supernovae}
\label{sec:calc}

We compute the expected number of strong lensing in future surveys in
the usual way, integrating the lensing cross section of each galaxy
over the galaxy population and redshift \citep[e.g.,][]{turner84}. As
source objects, we consider both QSOs and SNe, for which measurements
of time delays between multiple images will be possible due to their
time-variable nature. 

\subsection{The population of lensing galaxies}
\label{sec:lensgal}

An appropriate modeling of the lens galaxy population constitutes an
essential part of the prediction for lensing rates. We mostly follow
the model used in \citet{oguri08} to compute the expected number of
strong lensing by massive galaxies in future surveys. Throughout the
paper we conservatively consider only elliptical galaxies (E/S0) as
lensing objects; these likely comprise $\sim 80\%$ of the total
lensing probability
\citep{turner84,fukugita92,kochanek96,chae03,oguri06,moller07}. 
We assume that the mass distribution of each elliptical galaxy is
described by a Singular Isothermal Ellipsoid (SIE), whose convergence
is given by 
\begin{equation}
\kappa(x,y) = \frac{\theta_{\rm Ein}}{2}
\frac{\lambda(e)}{\sqrt{(1-e)^{-1}x^2+(1-e)y^2}},
\end{equation}
\begin{equation}
\theta_{\rm Ein} =  4\pi\left(\frac{v}{c}\right)^2\frac{D_{\rm
    ls}}{D_{\rm s}},
\label{eq:tein}
\end{equation}
where $v$ is the one-dimensional velocity dispersion of the lensing galaxy.
It has been shown that this model describe the mass distributions of
observed strong lensing elliptical galaxies very well
\citep{rusin01a,treu04,rusin05,koopmans06,koopmans09,gavazzi07}. The
parameter $\lambda(e)$, the so-called dynamical normalisation, is related
to the three-dimensional shape of lensing galaxies. Following
\citet{chae03}, we assume that there is an equal number of oblate and
prolate galaxies, and adopt the average value of the normalisations in
each of the oblate and prolate cases. 
As in \citet{oguri08}, we assume a Gaussian distribution for the
ellipticity $e$, with a mean of 0.3 and dispersion of 0.16. The
distribution is truncated at $e=0$ and $e=0.9$. 
 
In addition to the SIE model lens galaxy, we include external
shear to account for the effect of the lens environment
\citep[e.g.,][]{kochanek91,witt97,keeton97}. The lens potential
of external shear is given by 
\begin{equation}
\phi(x,y)=\frac{\gamma}{2}(x^2-y^2)\cos 2\theta_\gamma+\gamma xy\sin 
2\theta_\gamma.
\end{equation}
We assume the magnitude of the external shear, $\gamma$, follows a
log-normal distribution with mean 0.05 and dispersion 0.2~dex, 
which is broadly consistent with the level of external shear
expected from ray-tracing in $N$-body simulations
\citep{holder03,dalal04}. The orientation of the external shear,
$\theta_\gamma$, is assumed to be random. Although the external
convergence \citep[e.g.,][]{oguri05} is also vital particularly for
time delays, we do not include it explicitly in the calculation.
Instead, we will later consider its effects via an 
effective lens density profile slope,
which will be the subject of detailed discussion. 

For the velocity function of early-type galaxies, we adopt that
derived from the SDSS data \citep{choi07}, which is fitted by a
modified Schechter function of the following form:  
\begin{equation}
\frac{dn}{dv}=\phi_*\left(\frac{v}{v_*}\right)^\alpha
\exp\left[
  -\left(\frac{v}{v_*}\right)^\beta\right]\frac{\beta}{\Gamma 
(\alpha/\beta)}\frac{dv}{v}.
\label{eq:vf}
\end{equation}
Here, ($\phi_*$, $v_*$, $\alpha$, $\beta$)$=$($8.0\times
10^{-3}h^3{\rm Mpc^{-3}}$, $161\,{\rm km\,s^{-1}}$, $2.32$, $2.67$).
We do not include redshift evolution, i.e., we apply the mass
distribution and velocity function to galaxies at any redshifts.
Such a non-evolving  model has been shown to reproduce the abundance
of strong lenses in the latest radio \citep{chae07} and optical
\citep{oguri08} lens surveys, although see, e.g., \citet{mitchell05}
for an  analysis that takes evolution into account.

\subsection{Lensing rates}
\label{sec:lensprob}

The lensing probability for a given source at redshift $z_{\rm s}$ and with
luminosity~$L$ is given by
\begin{equation}
p=\int_{\theta_{\rm min}}^{\theta_{\rm max}} d\theta \int_0^{z_{\rm s}} 
dz_{\rm l} \frac{c\,dt}{dz_{\rm l}}(1+z_{\rm l})^3 \left[
\frac{dn}{dv}\frac{dv}{d\theta}\sigma_{\rm lens} \right]_{v=v(\theta)},
\label{eq:lensprob}
\end{equation}
where $\theta$ is the image separation between multiple
images,\footnote{For strong lens systems withe multiple images more
  than two, we define the image separation by the maximum separation
  between any pair of images.} $z_{\rm l}$ indicates the redshift of lensing
objects, and $dn/dv$ denotes the velocity function of lensing galaxies
given by equation (\ref{eq:vf}). Unless otherwise specified, we adopt
the minimum image separation of $\theta_{\rm min}=0.5''$ and maximum
image separation of $\theta_{\rm max}=4''$. Although a non-negligible
fraction of strong lenses have image separations larger than $4''$,
such lensing is caused by groups or clusters of galaxies rather than a
single massive galaxy \citep[e.g.,][]{oguri06}.
In this work we focus on galaxy-scale lenses (for which simple isothermal
density profiles are good models), for two reasons. Firstly, they do
dominate the lens abundance, which we will need to be high in order to
beat down the statistical errors on the cosmological parameters.
Secondly, the image positions are easier to fit with simple models,
reflecting the fact that massive galaxies are older, more relaxed
systems than groups and clusters. 
On low mass scales, 
lenses with image separation less than $0.5''$ were deemed to be too
difficult to measure, particularly in the ground-based surveys.
Indeed, when we consider the yields of specific surveys below, we take into
account the predicted image quality explicitly. 

The lensing cross section $\sigma_{\rm lens}$ is determined by the
structure of the lens potential. As discussed in \S\ref{sec:lensgal},
we adopt an SIE plus external shear as the lens potential of each
galaxy. To take account of the magnification bias, we compute the biased
cross section \citep{huterer05} as follows:
\begin{equation}
\sigma_{\rm lens}=\int
\frac{d\mathbf{u}}{\mu}\frac{d\Phi/dL(L/\mu)}{d\Phi/dL(L)},
\end{equation}
where $d\Phi/dL$ is the source luminosity function and $\mu$ is the
magnification factor. The integral is performed over the region where
multiple images are generated. In practice, we compute the biased cross
section numerically using Monte Carlo sampling of multiple images. We
solve the lens equation using the software package 
{\sc glafic} (M. Oguri, in preparation). We compute the biased cross
sections for double (two-images), quadruple (quad; four-images), and
naked cusp (three-images) lenses separately in order to study the
image multiplicity. For double lenses, we place a condition that the
flux ratio must be larger than $0.1$ and perform the integral over the
region that satisfies this condition. This is because asymmetric
double lenses with large flux differences are sometimes very difficult
to locate in observations as a result of dynamic range problems caused
by the brighter image. 

The choice of the magnification factor $\mu$ depends on the way in which
strong lens systems are identified in the survey data. Bearing in mind 
the promising variability selection techniques suggested by 
\citet{pindor05} and \citet{kochanek06b}, we adopt the magnification
factor of the fainter image for double lenses as $\mu$. For quad
and cusp lenses,  we adopt the magnification factor of the third-brightest
image, because the second-brightest image is sometimes located very
close to the brightest image if it is produced near the fold
catastrophe (where only one eigenvalue of magnification matrix is zero). 
The two images in a fold-configuration lens
may be hard to deblend because of their
small angular separation, which may result in failure to identify them
as strongly lensed.  Another reason is that the third brightest image
tends to arrive first, a feature we will return to later
(\S\ref{sec:delay}). 

Once the lensing probability for a given source is computed, it is
straightforward to calculate the expected strong lens abundance. We
obtain the differential number distribution of strong lenses by
integrating the lens probability over the source population:
\begin{equation}
\frac{dN}{dz_{\rm s}}=\int_{-\infty}^{M_{\rm max}} dM 
\frac{d\Phi}{dM}\frac{dV}{dz_{\rm s}}p,
\label{eq:dndzs}
\end{equation}
where $p$ is given by equation (\ref{eq:lensprob}) and $M$ denotes the
absolute magnitude of sources. The limiting absolute magnitude is
simply converted from the magnitude limit of the survey considered.
The volume factor $dV/dz_{\rm s}$ is given by
\begin{equation}
\frac{dV}{dz_{\rm s}}=\Omega D_{\rm s}^2\frac{c\,dt}{dz_{\rm s}}(1+z_{\rm s})^3,
\end{equation}
with $\Omega$ being the solid angle corresponding to the survey area.

Finally, the total number of strong lenses is
\begin{equation}
N=\int dz_{\rm s}\frac{dN}{dz_{\rm s}}.
\end{equation}
Note that we did not impose any restrictions on the luminosity of the
lensing galaxies. If a lens galaxy is much brighter than lensed
images, such lens systems may prove difficult to identify from the
imaging data alone. This effect becomes particularly important for
lensing of faint sources \citep[e.g.,][]{kochanek96,wyithe02}. 
However, since we are considering cadenced surveys where the 
strong lens search will make use of differenced survey images
\citep{kochanek06b}, we assume that those systems with bright lensing
galaxies can still be successfully identified because the lens galaxy
component (as well as any lensed host galaxy light) should be cleanly
subtracted. In practice, a bright lens galaxy will act as an
additional source of noise; we neglect this here, since many of the
surveys we consider are ground-based and will have background-limited
images.

\subsection{Number density of quasars}
\label{sec:qso}

We adopt the standard double power-law for the quasar luminosity
function. Specifically, we adopt the following parametric form:
\begin{equation}
\frac{d\Phi_{\rm QSO}}{dM}=
\frac{\Phi_*}{10^{0.4(\alpha+1)(M-M_*)}+10^{0.4(\beta+1)(M-M_*)}}.
\label{eq:lf_qso}
\end{equation}
Here $M$ refers to the absolute $i$-band magnitude of quasars.
We fix the faint end slope of $\beta=-1.45$ and the bright end slope
of $\alpha=-3.31$ which was obtained in the combined analysis of the
SDSS and 2dF \citep{richards05}. At $z>3$, however, we modify the
bright end slope and adopt the shallower slope of $\alpha=-2.58$ which
was suggested by observations \citep{fan01}. 

\begin{figure}
\begin{center}
 \includegraphics[width=0.95\hsize]{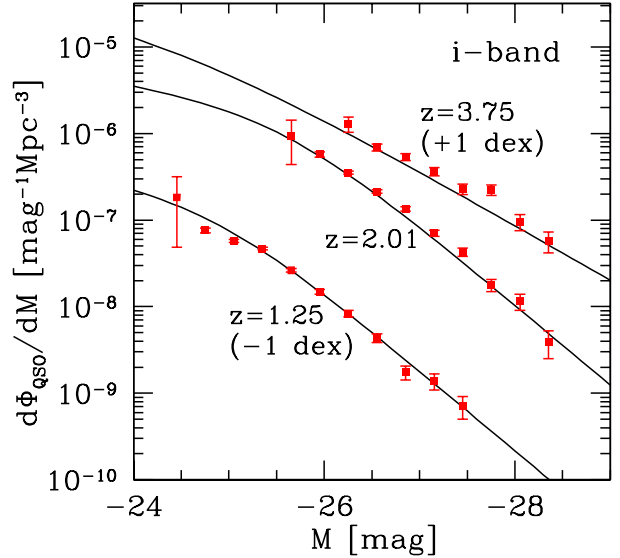}
\end{center}
\caption{Our model of the $i$-band quasar luminosity function
  (eq. [\ref{eq:lf_qso}]), compared with the observed quasar
  luminosity function from the SDSS DR3 \citep{richards06}. We present
  results at three different redshifts, $z=1.25$, $2.01$, and $3.75$, 
  in order to show that our model successfully reproduces the
  abundance of the SDSS quasars for a wide range of
  redshifts. Luminosity functions for $z=1.25$ ($z=3.75$) are shifted
  by $-1$~dex ($+1$~dex) to avoid overlap.
\label{fig:lf_qso}}
\end{figure}

In addition to this change of slope, we model the redshift evolution
of the luminosity function as pure luminosity evolution. Specifically,
the break absolute magnitude $M_*$ is described as
\begin{equation}
M_*=-20.90+5\log h-2.5\log f(z),
\end{equation}
\begin{equation}
f(z)=\frac{e^{\zeta z}(1+e^{\xi z_*})}{(\sqrt{e^{\xi z}}+\sqrt{e^{\xi z_*}})^2},
\end{equation}
where the zero-point value of $M_*$ , as well as the normalisation of
the luminosity function $\Phi_*=5.34\times10^{-6}h^3$~Mpc$^{-3}$, is
taken from \citet{richards05}. The parametric form of the redshift
evolution, $f(z)$, is taken from \citet{madau99} with some modifications to
achieve a better match to the observed evolution of the quasar
luminosity function. We then determine the values of the parameters
$\zeta$, $\xi$, and $z_*$ to reproduce the quasar luminosity function
from $z=0$ to $5$ obtained from the SDSS Data Release 3 \citep[DR3,
][]{richards06}. We find the best fit values to be ($\zeta$, $\xi$,
$z_*$)=(2.98, 4.05, 1.60), which we adopt throughout the paper. In
Figure~\ref{fig:lf_qso}, we compare our best-fit luminosity function
with the SDSS result of \citet{richards06}. 

In order to convert the $i$-band absolute magnitude to the apparent
magnitude, we need the appropriate K-correction. We adopt that
presented by \citet{richards06}. 

\subsection{Number density of supernovae}
\label{sec:sne}

The rate of occurrence of SNe is closely related to the star formation
rate. In this paper we adopt the cosmic star formation rate presented
by \citet{hopkins06},  
\begin{equation}
\rho_{\rm SFR}(z)=\frac{(0.0118+0.08z)h}{1+(z/3.3)^{5.2}}
[M_\odot{\rm yr^{-1}Mpc^{-3}}],
\end{equation}
which is essentially the best-fit to the observed star formation rates
assuming the initial mass function of \citet{baldry03}. 

Recent studies of the host galaxies of SNe Ia suggest that delay times
($t_{\rm D}$) of SNe Ia cannot be described by a single value. It has
sometimes been assumed that SNe Ia are drawn from two populations,
``prompt'' (tracing star formation rates) and ``old'' (tracing stellar
masses of host galaxies) SNe
\citep{scannapieco05,mannucci05,mannucci06,sullivan06b,aubourg08}. 
However, such a two-component model may fail to explain the observed low
high-redshift SNe Ia rates \citep[e.g.,][]{dahlen08}. It is naturally
expected that more realistic models have widely distributed $t_{\rm
  D}$, because theoretically the delay time should depend on various 
parameters such as the mass of the companion star and the metellicity
\citep[e.g.,][and references therein]{kobayashi09}. In this paper, we
use the result of \citet{totani08} who constrained the distribution of
$t_{\rm D}$ assuming a power-law:   
\begin{equation}
f(t_{\rm D})\propto t_{\rm D}^{-1.08}\;\;\; (t_{\rm D}>0.1{\rm Gyr}).
\end{equation}
Since they found no strong evidence for the presence of the ``prompt''
($t_{\rm D}<0.1{\rm Gyr}$) component, we do not consider it. 
We can then compute the SN Ia rate as
\begin{equation}
n_{\rm Ia}(z)=\eta C_{\rm Ia}\frac{\int_{0.1}^{t(z)}\rho_{\rm SFR}[z(t-t_{\rm D})]
 f(t_{\rm D})dt_{\rm D} }{\int_{0.1}^{t(z=0)}f(t_{\rm
    D})dt_{\rm D}},
\end{equation}
The factor $C_{\rm Ia}=0.032 M_\odot^{-1}$ can be computed from the
stellar mass range of $3M_\odot<M<8M_\odot$ for SNe Ia and the initial
mass function of \citet{baldry03}. We assume a canonical efficiency of 
$\eta=0.04$ \citep[see also][]{strigari05,hopkins06}. 

\begin{figure}
\begin{center}
 \includegraphics[width=0.95\hsize]{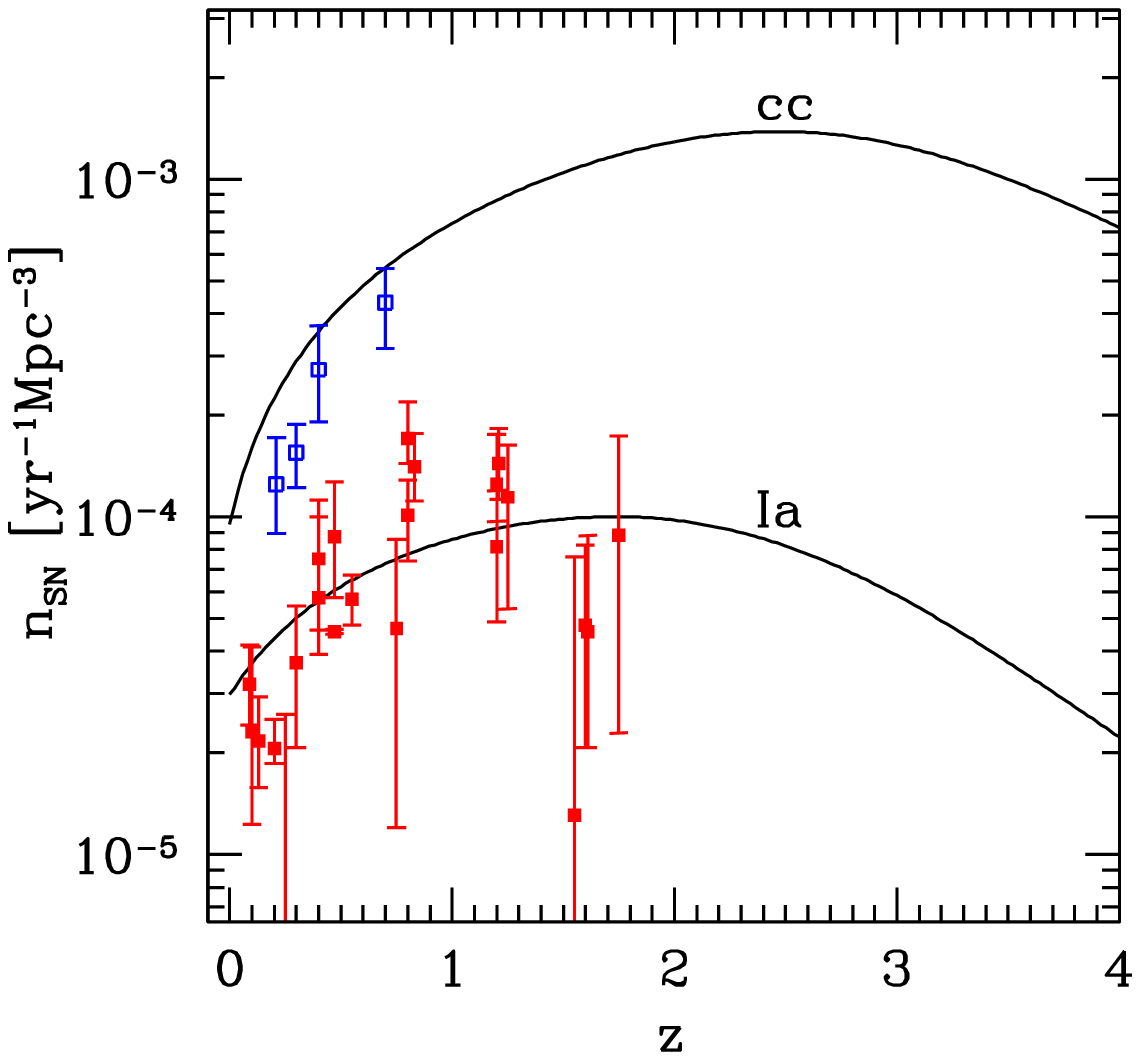}
\end{center}
\caption{The evolution of SN rates for type-Ia ({\it lower line}) and
  core-collapse (i.e., the sum of Ib/c, IIP, IIL, and IIn; 
  {\it upper line}) SNe adopted in this paper. See text for details of the
  model. Filled squares are recent measurements of SN Ia rates from
  \citet{hardin00}, \citet{pain02}, \citet{dahlen04}, \citet{blanc04}, 
  \citet{neill06}, \citet{poznanski07}, \citet{kuznetsova08}, 
  \citet{botticella08}, \citet{dilday08}, \citet{horesh08}, and 
  \citet{dahlen08}, whereas open squares are recent measurements of 
  core-collapse SN rates from \citet{dahlen04}, \citet{botticella08}, 
  and \citet{bazin09}. Errors indicate $1\sigma$ statistical errors, 
  and do not include any systematic errors. 
\label{fig:snrate}}
\end{figure}

On the other hand, core-collapse SNe are thought to be related
more directly to the star formation rate. They are associated with
the death of massive stars whose life time is significantly shorter
than the typical cosmological time scale, which suggests that the
SN rate of core-collapse SNe is simply proportional to
the cosmic star formation rate:
\begin{equation}
n_{\rm cc}(z)=C_{\rm cc} \rho_{\rm SFR}(z),
\end{equation}
where $C_{\rm cc}=0.0132 M_\odot^{-1}$ was derived assuming that
core-collapse SNe are produced in the stellar mass range
$8M_\odot<M<50M_\odot$, together with the initial mass function 
of \citet{baldry03}. Following \citet{oda05}, we adopt constant
relative proportions for four subclasses of core-collapse SNe.
Then SN rates are written as
\begin{equation}
n_{\rm X}(z)=f_{\rm X}n_{\rm cc}(z),
\end{equation}
where $f_{\rm Ib/c}=0.23$, $f_{\rm IIP}=0.30$, $f_{\rm IIL}=0.30$,
and $f_{\rm IIn}=0.02$ \citep{dahlen99}. Note that we do not consider
SN1987A-like SNe that are very faint compared with the other
core-collapse SNe. Recent analysis suggests that the fraction of
type-IIP SNe may actually be much higher \citep{smartt09}.

In Figure~\ref{fig:snrate}, we compare our model of SN rates with
recent measurements. We find that our model is reasonably consistent
with the observed SN rates, particularly given the additional systematic
errors they involve. The Figure indicates that our model of SN rates
is uncertain at the factor of $\sim 2$ level, suggesting that our predicted
numbers of lensed SNe will be similarly uncertain. In particular, 
our model appears to slightly over-predict the core-collapse SN rate,
though it is possible that current observations may miss a fraction of
core-collapse SNe given their wide range of intrinsic luminosities.
The uncertainty should be even larger at redshift $z\ga 2$ for type-Ia SNe,
and  $z\ga 1$ for core-collapse SNe, where no measurements of SN rates
have been obtained.  

We need not only SN rates, but also the brightness distributions of
SNe in order to make predictions of the lensed SNe abundance. In what
follows the magnitude of SNe refers to the peak magnitude, i.e., the
magnitude when the SN is brightest. Again following \citet{oda05}, we
assume the absolute magnitudes of SNe are Gaussian-distributed
\citep[see also][]{yasuda10}.  With this assumption the luminosity
function (in terms of $B$-band absolute magnitude $M$) can be written as
\begin{equation}
\frac{d\Phi_{\rm X}}{dM}=\frac{n_{\rm
    X}(z)}{(1+z)}\frac{1}{\sqrt{2\pi}\sigma_{\rm X}}
 \exp\left[-\frac{(M-M_{\rm X}^*)^2}{2\sigma_{\rm x}^2}\right],
\label{eq:lf_sn}
\end{equation}
where ($M_{\rm Ia}^*$, $M_{\rm Ib/c}^*$, $M_{\rm IIP}^*$, $M_{\rm
  IIL}^*$, $M_{\rm IIn}^*$)=($-19.06$, $-17.64$, $-16.60$, $-17.63$,
$-18.75$) (for $h=0.72$) and ($\sigma_{\rm Ia}$, $\sigma_{\rm Ib/c}$,
$\sigma_{\rm IIP}$, $\sigma_{\rm IIL}$, $\sigma_{\rm IIn}$)=(0.56,
1.39, 1.12, 0.90, 0.92) \citep{richardson02}. Note that the luminosity
function of SNe (eq. [\ref{eq:lf_sn}]) differs from that of QSOs
(eq. [\ref{eq:lf_qso}]) in that the former is in fact the number rate
(number per unit time). Thus the factor of $(1+z)^{-1}$ is introduced
to account for the cosmological time dilation.

We convert $B$-band absolute magnitudes to apparent magnitudes in the {\it
i}-band  by computing K-corrections using various SN template spectra. We
adopt the spectra at the peak presented by \citet{nugent02} for Ia,
\citet{levan05} for Ib/c, and \citet{gilliland99} for IIP, IIL, and IIn.

\section{Strong lenses in various surveys}
\label{sec:survey}

In this section, we predict the number of strongly lensed QSOs and SNe
for a selection of ongoing and planned surveys, 
using the model described in detail in
\S\ref{sec:calc}. 

\subsection{The expected number of lenses as a function of survey depth}
\label{sec:mlim}

\begin{figure}
\begin{center}
 \includegraphics[width=0.95\hsize]{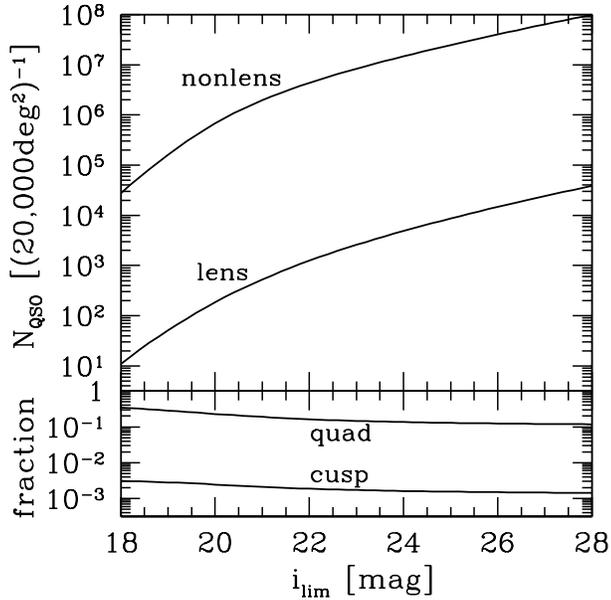}
\end{center}
\caption{The expected number of lensed QSOs as a function of the
  $i$-band limiting magnitude $i_{\rm lim}$. A fiducial survey area of
  $\Omega=20,000$~${\rm deg^2}$ is assumed. The number of non-lensed
  QSOs is also shown for reference. The lower panel shows the ratio of
  the number of quad (four-images) or naked cusp (three-images) lenses
  to the total number of lenses, as a function of $i_{\rm lim}$.   
\label{fig:qso_mlim}}
\end{figure}

First, we explore how the number of lenses detected increases with
survey depth.  In Figure~\ref{fig:qso_mlim}, we plot the number of
lensed QSOs (in a half-sky survey) as a function of $i$-band limiting
magnitude $i_{\rm lim}$.  The slope of these number counts is fairly
shallow, particularly at $i_{\rm lim}\ga 21$, which suggests that the
survey area is much more important than the survey depth when trying
to discover many strongly lensed QSOs. The lensing rate is $\sim
10^{-3.5}$, and does not depend very much on $i_{\rm lim}$ due to
the conflicting effects of increasing mean QSO redshift and
decreasing magnification bias. We note that the lensing rate is lower
than observed in the Cosmic Lens All-Sky Survey (CLASS), $\sim
10^{-2.8}$ \citep{browne03}, presumably because of the quite different 
magnification bias it involves (for instance, in the CLASS the total
magnification factor is used for the magnification bias). The recent
optical lens survey, the SDSS Quasar Lens Search (SQLS), has obtained
a lensing rate of $\sim 10^{-3.3}$, which is more consistent with the
calculation above \citep[see][]{inada08}. 
The fraction of quad lenses decreases
from $\sim 30\%$ for $i_{\rm lim}=18$ to $\sim 10\%$ for $i_{\rm
  lim}=28$, which is roughly consistent with previous calculations
\citep{rusin01b,huterer05,oguri07b}. The small fraction ($\sim
10^{-3}$) of naked cusp lenses indicates that only very wide-field
surveys will be able to locate such rare image configurations. 
We again note
that our calculation is applicable only to galaxy scale lenses; 
naked cusp lenses are much more common at cluster scales, where the
radial density profiles of the lenses are shallower
\citep{oguri04,minor08}. We also note that our model lens galaxies
are all spheroids, with correspondingly low ellipticity. The lower
mass, disky lenses, that will make up a small minority ($\la 20\%$, 
see~\S\ref{sec:lensgal}) of any survey's yield, may be expected to be
more elongated and hence present a higher naked cusp fraction
\citep{keeton98}. 
 
\begin{figure}
\begin{center}
 \includegraphics[width=0.95\hsize]{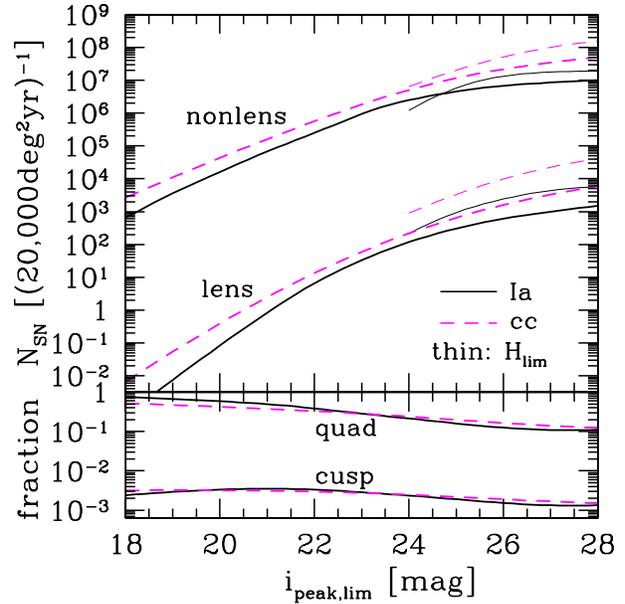}
\end{center}
\caption{The expected number of lensed SNe as a function of the
  $i$-band limiting peak magnitude, $i_{\rm peak,lim}$, for one
  year monitoring of a fiducial $\Omega=20,000$~${\rm deg^2}$ survey
  area. Upper curves indicate numbers of normal unlensed SNe, for
  reference.  Solid lines show expected numbers of SNe Ia, whereas
  dashed lines show the numbers of core-collapse SNe. The thin lines
  at $>24$~mag show the expected numbers of SNe as a function of
  $H_{\rm peak,lim}$ rather than $i_{\rm peak,lim}$, as relevant for 
  a near-infrared optimised space-based SN survey. The lower panel
  shows the fractions of quad and naked cusp lenses. 
\label{fig:sn_mlim}}
\end{figure}

In Figure~\ref{fig:sn_mlim} we plot the expected numbers of lensed
SNe. The slope of the number counts is steeper than it was in the case
of the lensed QSOs, suggesting that the survey depth is more important
for lensed SNe than it is for lensed quasars.
Lensed core-collapse SNe are more abundant than lensed
Ia by a factor of several. We find that the fraction of
quad systems changes very rapidly as a function of the
limiting magnitude. For shallow SN surveys, with, say, limiting peak
magnitude $i_{\rm peak,lim}\la 22$, the strong lens sample would be
approximately 50\% quads. The high quad fraction is due to very large
magnification bias inherent to such bright lenses. For deeper surveys
of $i_{\rm peak,lim}\ga 26$, however, the fraction of quad lenses
takes a typical value of $10-20\%$. The fraction of cusp lenses
remains low, $\sim 10^{-3.5}$. 

The number of (lensed) SNe does not increase very much beyond $i_{\rm
  peak,lim}\sim 24$. This is mainly because of the increasing
values of the K-correction at high-redshifts. Since SN spectra have UV
cutoff at $\sim 3000$~\AA, the $i$-band K-correction rises quickly at
$z\ga 1.7$, which significantly degrades the detectability of such
high-redshift SNe in $i$-band images (this effect has long been a
source of frustration for ground-based supernova cosmology projects).
To illustrate this point, in Figure~\ref{fig:sn_mlim} we also plot the
number of SNe as a function of $H$-band limiting peak magnitude,
$H_{\rm peak,lim}$. It is clear that the expected number of lensed SNe
is much larger in $H$-band than $i$-band for deep monitoring. The
steeper number counts in $H$-band indicates that the effect of
K-correction is much less important, illustrating the importance of
near-infrared imaging for detecting high-redshift SNe (lensed or
otherwise). 

\subsection{The expected number of lenses in individual surveys}
\label{sec:surveys}

\begin{table*}
 \caption{Properties of various time-domain surveys. These are fiducial 
  numbers based on assumptions we describe in the text; given are survey
  area $\Omega$, $10\sigma$ point source limiting magnitude $i_{\rm lim}$
  for one visit, one year, and for the final stacked survey image,
  corresponding assumed median image quality $\theta_{\rm PSF}$
  (seeing FWHM, in units of arcseconds), cadence (in days),  
  season length $t_{\rm season}$ (in months), survey length $t_{\rm survey}$
  (in years) and the effective survey duration 
   $t_{\rm eff} = t_{\rm survey} \cdot t_{\rm season}/12$ 
  (also in years). \label{table:surveys}} 
 \begin{tabular}{@{}lccccccccc}
  \hline
Survey & $\Omega$  & $i_{\rm lim}$(visit)
    & $i_{\rm lim}$(year) & $i_{\rm lim}$(total)  
    & $\theta_{\rm PSF}$ & cadence & $t_{\rm season}$  
    & $t_{\rm survey}$ & $t_{\rm eff}$  \\                
    & [deg$^2$] &  &  &  & [arcsec] &  [days] & [months]
    & [years] & [years] \\                
 \hline
SDSS-II    & 300   & 21.3 & 22.9      & 23.5 & 1.4  & 5  & 3 & 3  & 0.75 \\
SNLS       & 4     & 24.3 & 26.3      & 27.1 & 0.9  & 4  & 5 & 5  & 2.1 \\ 
PS1/$3\pi$ & 30000 & 21.4 & 22.2      & 22.7 & 1.0  & 30 & 2 & 3  & 0.5 \\
PS1/MDS    & 70    & 23.3 & 25.2      & 25.8 & 1.0  & 4  & 6 & 3  & 1.5 \\
DES/wide   & 5000  & 23.6 & 24.0      & 25.1 & 0.9  & 30 & 1 & 5  & 0.4 \\
DES/deep   & 6     & 24.6 & 26.1      & 27.0 & 0.9  & 5  & 3 & 5  & 1.3 \\
HSC/wide   & 1500  & 24.9 & 24.9      & 25.5 & 0.75 & $\cdots^a$& 0 & 3  & 0.0 \\
HSC/deep   & 30    & 25.3 & 26.6      & 26.6 & 0.75 & 5  & 2 & 1  & 0.2 \\
JDEM/SNAP  & 15    & 27.1 & $\cdots^b$& 29.7 & 0.14 & 4  & 12& 1.3& 1.3 \\
LSST       & 20000 & 23.3 & 24.9      & 26.2 & 0.75 & 5  & 3 & 10 & 2.5 \\
 \hline
 \end{tabular}
\newline\footnotesize
\flushleft{$^a$The HSC/wide survey is zero cadence: with 
only one observing epoch 
per year it would not be suitable for useful lensed
SN detection.}\vspace{-0.5\baselineskip}
\flushleft{$^b$The JDEM/SNAP survey is assumed to be undertaken as 
a single 1.3~year observing campaign, so does not offer the yearly
stack detection strategy. In practice lensed quasars would be
detectable simply from the resolved image geometries in any case.}
\end{table*}
%

\begin{table*}
 \caption{The expected numbers of lensed QSOs in various
   time-domain surveys. The numbers are either for detections only,
   using stacked images within each survey year, or for well-measured 
   time delays using each visit's image. Some wide-field surveys do
   not have high enough cadence to measure time delays, and therefore
   those surveys are useful only for detecting lensed QSOs (see text
   for more details). We adopt the minimum image
   separation $\theta_{\rm min}=(2/3)\theta_{\rm PSF}$ for all surveys. The
   numbers of non-lensed QSOs detectable in the surveys are also
   shown for reference. Percentages in parentheses indicate the 
  fraction of quad lenses. \label{table:lqsos}} 
 \begin{tabular}{@{}lccccc}
  \hline
& \multicolumn{2}{c}{QSO (detected)} & 
&\multicolumn{2}{c}{QSO (measured)} \\
\cline{2-3}  \cline{5-6}  
Survey & $N_{\rm nonlens}$ & $N_{\rm lens}$ &      
       & $N_{\rm nonlens}$ & $N_{\rm lens}$ \\                
 \hline
SDSS-II   
    & $1.18\times 10^5$ &  26.3 (15\%) &
    & $3.82\times 10^4$ &   7.6 (18\%) \\
SNLS      
    & $9.23\times 10^3$ &  3.2 (12\%) &
    & $3.45\times 10^3$ &  1.1 (13\%) \\
PS1/$3\pi$
    & $7.52\times 10^6$ &  1963 (16\%) &
    & $\cdots$ &  $\cdots$ \\
PS1/MDS   
    & $9.55\times 10^4$ &  30.3 (13\%) &
    & $3.49\times 10^4$ &   9.9 (14\%) \\
DES/wide  
    & $3.68\times 10^6$ &  1146 (14\%) &
    & $\cdots$ & $\cdots$ \\
DES/deep  
    & $1.26\times 10^4$ &   4.4 (12\%) &
    & $6.05\times 10^3$ &   2.0 (13\%) \\
HSC/wide  
    & $1.76\times 10^6$ &   614 (13\%) &
    & $\cdots$  &  $\cdots$ \\
HSC/deep  
    & $7.96\times 10^4$ &  29.7 (12\%) &
    & $4.30\times 10^4$ &  15.3 (13\%) \\
JDEM/SNAP 
    & $5.00\times 10^4$ &  21.8 (12\%) &
    & $5.00\times 10^4$ &  21.8 (12\%) \\
LSST      
    & $2.35\times 10^7$ &  8191 (13\%) &
    & $9.97\times 10^6$ &  3150 (14\%) \\
 \hline
 \end{tabular}
\end{table*}
%

\begin{table*}
 \caption{The expected number of detected lensed SNe (type Ia
  and core collapse) in various time-domain surveys. We adopt the minimum image
  separation $\theta_{\rm min}=(2/3)\theta_{\rm PSF}$ for all surveys.
  The numbers of non-lensed sources detectable in the surveys are also
  shown for reference. Percentages in parentheses indicate the
  fraction of quad lenses. For lensed SNe, we adopt the peak
  magnitude limit of $i_{\rm peak,lim}=i_{\rm lim}-0.7$ in actual
  calculations so that the lightcurves of lensed SN images can well be
  traced.\label{table:lsne}} 
  \begin{tabular}{@{}lcccccc}
  \hline
& \multicolumn{2}{c}{SN (Ia)} & &  \multicolumn{2}{c}{SN (cc)} &\\
\cline{2-3}  \cline{5-6} 
Survey & $N_{\rm nonlens}$ & $N_{\rm lens}$ &                
       & $N_{\rm nonlens}$ & $N_{\rm lens}$ &
    Note \\
 \hline
SDSS-II  
    & $4.34\times 10^2$ &  0.003 (54\%) &
    & $1.09\times 10^3$ &  0.01  (40\%) & \\
SNLS     
    & $7.52\times 10^2$ &  0.03  (24\%) &
    & $1.44\times 10^3$ &  0.05  (26\%) & \\
PS1/$3\pi$
    & $3.34\times 10^4$ &  0.28  (53\%) &
    & $8.23\times 10^4$ &  0.97  (39\%) & detections only\\
PS1/MDS   
    & $2.93\times 10^3$ &  0.09 (32\%) &
    & $6.05\times 10^3$ &  0.16 (30\%) & \\
DES/wide  
    & $8.30\times 10^4$ &   2.7 (29\%) &
    & $1.62\times 10^5$ &   4.9 (29\%) & detections only\\
DES/deep  
    & $8.95\times 10^2$ &  0.04 (22\%) &
    & $1.80\times 10^3$ &  0.07 (24\%) & \\
HSC/deep  
    & $1.10\times 10^3$ &  0.06 (18\%) &
    & $2.56\times 10^3$ &  0.13 (21\%) & \\
JDEM/SNAP$^a$
    & $1.36\times 10^4$ &  2.9  (13\%) &
    & $5.39\times 10^4$ &  12.0 (18\%) & \\
LSST      
    & $1.39\times 10^6$ &  45.7 (32\%) &
    & $2.88\times 10^6$ &  83.9 (30\%) & \\
 \hline
 \end{tabular}
\footnotesize
\flushleft{$^a$ Instead of the $i$-band, we adopt an 
    $H$-band magnitude limit
    of $H_{\rm lim}=26.8$ to predicted the number of (lensed) SNe,
    since in practice the detection in space will be done in the 
    near-infrared to optimise the number of high-redshift SN sources.}
\end{table*}
%

Next we consider strongly lensed QSOs and SNe in several specific
time-domain surveys. Each survey is characterised by the survey area
($\Omega$) and AB magnitude limit $i_{\rm lim}$, defined by the $i$-band
$10\sigma$ detection limit for point sources. In some
cases, we estimated an approximate $10\sigma$ limit from the
$5\sigma$ detection limit quoted in the reference 
by subtracting 0.7~mag. In some cases  we
also made rough conversions between image qualities, in order to
represent accurately the conditions currently expected to be 
achievable. These approximations
were confirmed to be accurate at the 5\% level using various exposure
time calculation  tools.\footnote{Specifically, we used the CFHT
  MegaCam facility ``DIET,'' \\ 
\texttt{http://rpm.cfht.hawaii.edu/\~{}megacam/diet/DIET.rpm},\\
and also the LSST ETC at\\ 
\texttt{http://dls.physics.ucdavis.edu:8080/etc4\_3work/\\
servlets/LsstEtc.html}}
For lensed SNe the interpretation of the magnitude limit is not obvious
because the brightnesses of SNe change drastically with time. We set
the condition that (lensed) SNe will be detected if the peak magnitude
is more than 0.7~mag brighter than the magnitude limit of each
visit. This requirement is chosen to ensure that the lightcurve of
each SN is traced reasonably well, and is consistent with the
$25\sigma$ peak brightness detection required by \citet{young08}. In
addition, for lensed SNe we need to specify the effective survey
duration $t_{\rm eff}$ to predict the total number of lensed
SNe. Depending on the survey duration and design, a fraction of lensed
SN events may be missed simply because one of the multiple images
appears outside the survey time; in this paper we do not consider this
``time delay bias'' \citep{oguri03a} for simplicity. We also adopt
different minimum image separation $\theta_{\rm min}$ for different
surveys, depending on their typical seeing sizes. Specifically we
assume $\theta_{\rm min}=(2/3)\theta_{\rm PSF}$, where 
$\theta_{\rm PSF}$ is a typical seeing FWHM of each survey, which is
empirically consistent with with previous optical lens surveys
\citep[e.g.,][]{inada08}. 

Finally, we note that detecting a time-variable lens, and being able
to measure its time delay, are two very different things. Since
quasars are marked out by their long term variability
\citep[e.g.,][Schmidt et al, in preparation]{kelly09}, a good strategy
for detecting lensed quasars may be to stack images within each survey
year, and then look for variability between years. For this reason we
consider the limiting magnitude of a survey per year's observation for
detecting quasar lenses. Measuring time delays requires dense
sampling: we assume that a cadence of $<10$ days and a season length
of $\geq 3$~months are the minimum needed to allow time delay
estimation, based on results of previous lens monitoring campaigns
\citep[e.g.][]{fassnacht02}. We note that quantifying time delay
accuracy as a function of photometric precision, season length, and
cadence would make a very valuable future study. Detecting {\it
  useful} lensed SNe will also require high cadence survey data, since
for these lenses there is no option of building up time delay
precision in subsequent seasons. 

We now briefly describe ongoing and future time-domain surveys which
we consider in this paper. Survey parameters for these surveys,
including survey areas, limiting magnitudes, and survey durations,
are listed in Table~\ref{table:surveys}.

\begin{itemize}
\item Sloan Digital Sky Survey-II Supernova Survey (SDSS-II SN). This
  survey was designed to locate relatively low-redshift ($z \sim
  0.05-0.35$) supernovae in the so-called ``stripe 82'' region of the
  SDSS \citep{frieman08}. It involved monitoring  an area of
  300~deg$^2$ with five broad-band filters; the limiting magnitude at
  each epoch is $i_{\rm lim}=21.3$. The survey consisted of three
  three-month observing seasons, corresponding to an effective survey
  duration for the SN search of $t_{\rm eff}=0.75$~yr. We assume a
  typical seeing of $1\farcs4$, and a monitoring cadence of 5 days, 
  giving 18 visits per field per year.\\

\item Supernova Legacy Survey (SNLS). Running for 5 years, with
  roughly five-month observing seasons for a total effective survey
  duration of 2.1 years, the SNLS is the cadenced part of the CFHT
  legacy survey, designed for discovery, classification and monitoring 
  of supernovae at intermediate redshift. The much larger ``Wide''
  part of the survey has only static images that do not
  (straightforwardly) enable a time delay lens search. The ``Deep'' 
  fields cover 4~deg$^2$. In calculating the $i$-band magnitude limit
  $i_{\rm lim}=24.3$, we assumed (constant) median seeing $0\farcs9$, and
  1800s exposure time per epoch~\citep{sullivan06a}. We also assume a 4-day 
  cadence, leading to 38 visits per field per year.\\ 

\item Pan-Starrs 1 (PS1). The first 1.8-m telescope of the Pan-STARRS
  project is carrying out two surveys, the Medium Deep Survey (MDS)
  and the $3\pi$ survey. Both are cadenced, but at different
  rates. The MDS covers 70~deg$^2$, and we assume that each of its
  fields will be observed for three six-month seasons with high
  (typically 4-day) cadence. We anticipate the $3\pi$ survey covering
  30,000~deg$^2$ at lower (typically one~month)  cadence, and assume three
  three-month seasons \citep[we follow][in anticipating using all the
  filters in the SN identification]{young08}. We conservatively assume a
  median image quality of $1\farcs0$, and use this to estimate
  $10\sigma$ $i$-band single-visit  limiting magnitudes for each
  survey of 23.3 (MDS) and 21.4 ($3\pi$), extrapolating approximately 
  from the figures given in the design reference mission~\citep{PS1}. \\ 

\item Dark Energy Survey (DES). To be carried out with a new
  wide-field camera installed on the Blanco 4-m telescope at CTIO, 
  this survey, to start in 2012, is also planned to have two
  components, a wide and a deep \citep{DES}. As with PS1, only the
  deep fields will have sufficient cadence for time delay estimation,
  but the wide fields will again allow detection of time-delay lenses
  by their variability. While the details of the surveys are still
  being decided, we adopt the following fiducial estimates: a deep
  survey covering 6~deg$^2$ for five years (in nearly three-month
  seasons, with 5-day cadence, giving 16 visits per year),  and a
  wide survey covering 5000~deg$^2$ for five years, with two observing
  epochs per year \citep{frieman09}. We take the effective season
  length of the wide survey to be one month, in order to estimate the
  number of lensed supernovae seen: this is about the period over
  which 2 images might be simultaneously visible. 
  We assume $i_{\rm lim}=24.6$ per epoch in  
  the deep fields, and  $i_{\rm lim}=23.6$ per epoch in the wide fields
  (J.~Annis, priv.\ comm.).\\

\item Hyper Suprime-Cam (HSC). The HSC is a next-generation
  prime-focus camera for the Subaru 8.2-meter telescope. The
  field-of-view is 1.5~degree in diameter, and the expected image
  quality is about $0\farcs75$ across the entire field-of-view. The
  survey plan with the HSC is not yet finalised and still very
  uncertain, but one of the three survey components (``deep'')  has
  both area and depth enough to locate many high-redshift
  supernovae. Here we tentatively assume its survey area to be
  30~deg$^2$ and the limiting magnitude to be $i_{\rm lim}=25.3$, with a
  single two~month season (which therefore may not be sufficient to
  measure time delays between multiple images). At an assumed cadence
  of 5 days, each field would be observed 12 times. Another component
  (the ``wide'')  may cover 1500~deg$^2$, with the imaging again being
  built up one exposure per year, for two or three years. As with
  PS1/$3\pi$ and  DES/wide, we flag this survey as useful for
  detection but not time delay measurement. We do not consider
  the ``ultra-deep'' survey here as its survey area will be rather small.\\ 

\item Joint Dark Energy Mission (JDEM). The design of this space-based
  observatory is still very  uncertain, but we can assume that if such
  a facility is built with high redshift supernovae in mind, it will
  have an infrared-optimised imager.  In this work we compute, as an
  example, predictions for a JDEM that has indeed been optimised for
  supernova work, and assume the specifications of the
  Supernova/Acceleration Probe (SNAP) as our fiducial telescope
  \citep{aldering04}: we assume a 15~deg$^2$ sky area observed to a
  depth of $i_{\rm lim}=27.1$ and $H_{\rm lim} = 26.8$ per epoch at
  high cadence (4 days)  for 1.3 years (where we have again corrected
  to get $10\sigma$ limits). Note that we only consider in this work
  the SN survey of this strawman JDEM; a weak lensing survey in space
  should find large numbers of lensed quasars even if the imaging is
  not cadenced at all, as the single epoch high resolution imaging
  alone will allow identification of multiple image systems 
  \citep[see, e.g.,][for some estimates]{kuhlen04}. However, time delay
  measurements would be possible only in the SN fields. \\ 

\item Large Synoptic Survey Telescope (LSST). This 6.7-m effective
  aperture dedicated survey telescope is designed to survey
  20,000~deg$^2$ of sky for ten years. The ``universal cadence''
  survey outlined by \citet{ivezic08} sees each field being observed
  in 6 filters at typically 5-day cadence in the $r$ and $i$-bands.
  A typical field's observing season would last three months or more
  (with the redder filters sampling a longer season), giving
  some 20 exposures per field per filter per year. We estimate that
  the $10\sigma$ limiting magnitude in $0\farcs75$ seeing will be
  around 23.3, and adopt this as our fiducial single-epoch depth. 
  \citeauthor{ivezic08} note that ``mini surveys'' could be carried
  out with LSST on smaller sky patches to greater depth; we do not
  consider any such survey here, but note that it would provide
  higher accuracy time delays on an inevitably smaller number of
  lenses. \\  
\end{itemize}

Tables~\ref{table:lqsos} and \ref{table:lsne} summarises the expected
numbers of lensed QSOs and SNe in each survey. We find that the LSST
will find the largest number of lenses with measureable time delays,
for both quasars and SNe. 
The wide surveys of the HSC, DES and PS1 projects will also discover large
numbers of  lensed quasars thanks to their wide field-of-view, 
but they will not have sufficient time domain sampling to measure the time
delays between quasar images. The corresponding deep surveys, or a
spaced-based JDEM/SNAP survey, will provide time delay measurements, but
only for $\sim 60$ lensed quasars, because of their small combined
field-of-view.
The LSST is unique because it has {\it both} the wide field-of-view to
cover many quasars and the frequent time domain sampling for
monitoring of lens systems, allowing us to obtain time delay
measurements for $\sim 3000$ lensed quasars. It will be quite challenging
to find lensed SNe in ground-based surveys before LSST, 
as the expected number of
lensed SNe discovered is only $\sim 1$. In contrast, the very deep
monitoring by a JDEM/SNAP survey would allow us to locate $\sim 15$
lensed SNe in total, despite the relatively small field of view of
$15$~deg$^2$. The LSST will also be powerful for detecting lensed SNe; our
calculation indicates that the LSST will discover $\sim 130$ lensed
supernovae during its 10-year survey.

\section{Generating Mock Catalogues: A Monte-Carlo approach}
\label{sec:mock}

Here we compute the expected number of strong lenses, and their
distributions with lens and source property, in a different way from
the standard technique described in \S\ref{sec:calc}.  Essentially it
is a semi-analytic technique based on Monte-Carlo realisations of lens
and source populations, and is an extension of the method used in
\citet{oguri09} to predict the all-sky distribution of large Einstein
radii produced by massive clusters. In essence, we perform the
necessary integrals by Monte-Carlo integration, and keep the sample
points drawn from the target distribution.  The goal is to produce
mock catalogues of lenses for a given survey, which can then be used to
investigate the feasibility of various science projects. 

\subsection{Technique description}
\label{sec:mccalc}

The specific procedure is as follows. First we randomly generate a
list of sources in a given survey area to a given depth, according to
the adopted source luminosity function. To allow for the magnification
by strong lensing, the limiting magnitude of the source catalogue needs
to be deeper than the actual magnitude limit of the survey. We adopt
$3-3.5$~mag deeper limit than the survey magnitude limit for
generating the source catalogue. This is sufficient, particularly given
our use of the fainter image magnification factor in computing the
magnification bias. Each source realisation contains information on
the redshift and apparent magnitude (without lensing magnification)
only. Since we ignore any spatial correlations between sources, the
whole source population is characterised just by a surface number
density. We then randomly generate a list of lenses according to the
model described in \S\ref{sec:lensgal}. For each mock lens galaxy, we
consider the rectangular region with an area of $\left[8\theta_{\rm
    Ein}({\rm max})\right]^2$, where $\theta_{\rm Ein}({\rm max})$ is
given by equation (\ref{eq:tein}) with $D_{\rm ls}/D_{\rm s}=1$,
randomly distribute source objects using the pre-generated catalogue,
and for each source solve the lens equation to see if it is
multiply-imaged. Again, the lens equation is solved using the software
{\it glafic} (M. Oguri, in preparation). For multiply-imaged sources,
we check whether they satisfy the detection criteria (the survey limiting
magnitude, image separation, and flux ratio), and record only lens
systems which survive these criteria. In this way we can generate mock
catalogues of lensed sources for a particular survey.

\begin{figure}
\begin{center}
 \includegraphics[width=0.95\hsize]{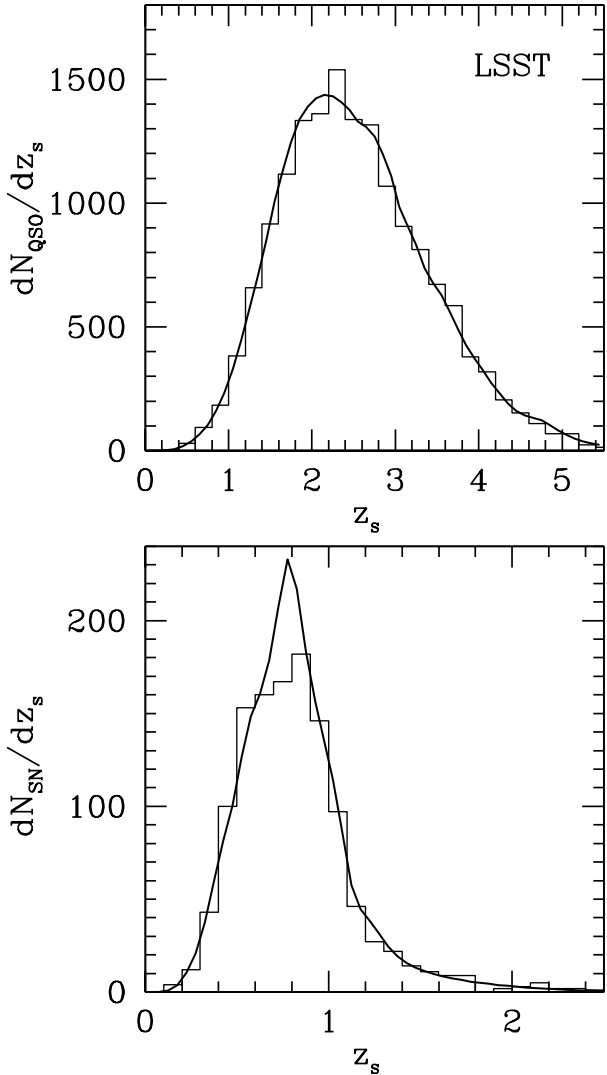}
\end{center}
\caption{The predicted redshift distributions of lensed QSOs
  ({\it upper panel}) and SNe ({\it lower panel}). 
  The histograms are distributions
  calculated from the mock lens catalogue constructed using the
  Monte-Carlo technique, whereas the solid curves are analytic
  results from equation (\ref{eq:dndzs}).
\label{fig:zsdist}}
\end{figure}

\begin{figure}
\begin{center}
 \includegraphics[width=0.95\hsize]{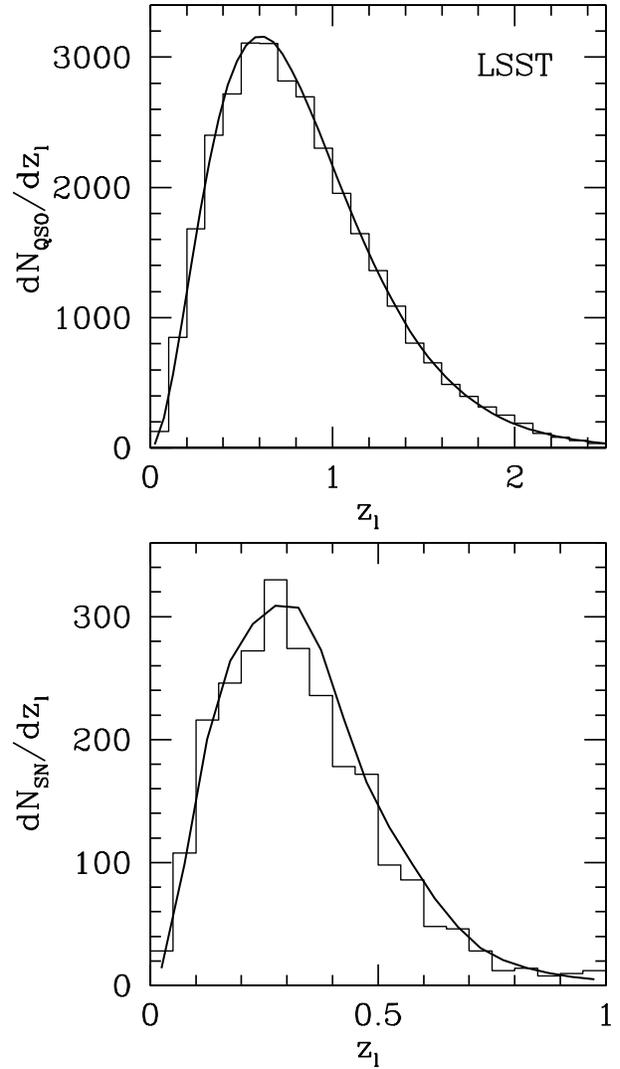}
\end{center}
\caption{Same as Figure~\ref{fig:zsdist}, but the redshift
  distributions of lensing galaxies are shown.
\label{fig:zldist}}
\end{figure}

Although the new Monte-Carlo technique tends to be more time-consuming 
than the traditional analytic approach, it has several important
advantages. (1) In this approach we can add more parameters to the
lens model with little additional computational cost. In the
traditional analytic approach, adding additional lens parameters
requires an additional integral, particularly if the distributions of
the parameters are not independent of each other.  This exponential
growth in run time with model complexity is in marked contrast with
the Monte-Carlo approach, in which the CPU time required scales
approximately linearly with the number of model parameters.
(2) It is straightforward to include even quite complicated 
lens selection functions, e.g., those dependent on many observables
such as flux ratios, image separations, time delays, and the
properties of lensing galaxies. (3) From the resulting mock catalogue we
can examine the expected distributions of lens and source parameters
quite easily. This is important because the population of strong
lenses generally differs from the population of general lens and
source populations
\citep[e.g.,][]{oguri05,moller07,rozo07,mandelbaum09}, and  
yet we would very much like to use the former to understand the latter.

\subsection{Mock lens catalogue}
\label{sec:catalog}

\begin{figure}
\begin{center}
 \includegraphics[width=0.95\hsize]{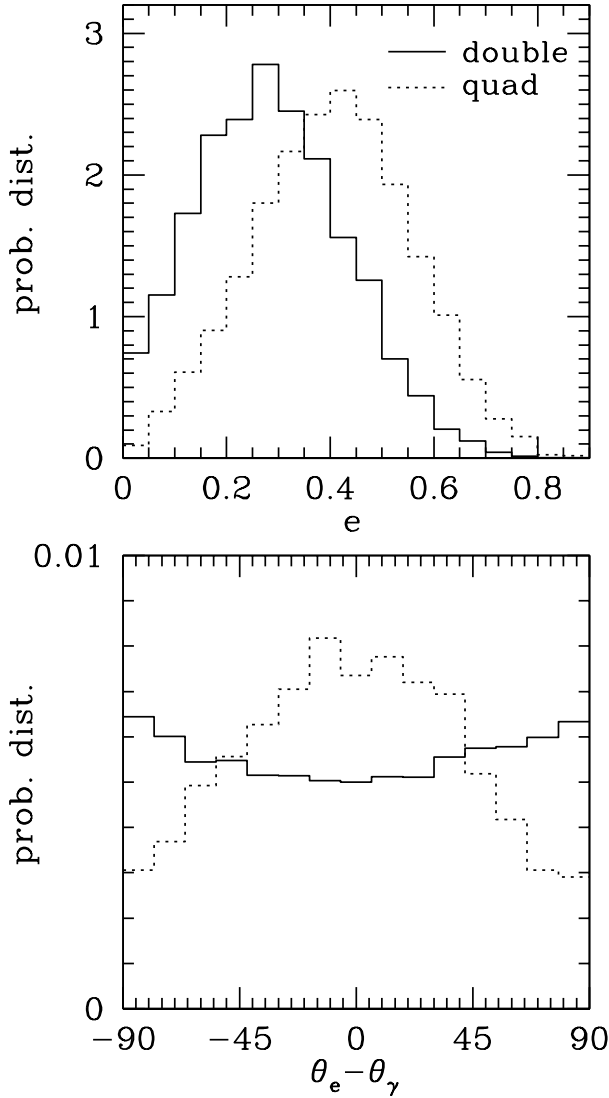}
\end{center}
\caption{{\it Upper panel:} 
  The probability distribution of the lens galaxy
  ellipticity $e$ for double ({\it solid}) and quad ({\it dotted})
  lenses, obtained from the mock lens catalogue. 
  {\it Lower panel:} The
  probability distribution of $\theta_e-\theta_\gamma$, where
  $\theta_e$ is the position angle of the lensing galaxy and
  $\theta_\gamma$ is the position angle of external shear. Since we
  assumed random distributions for $\theta_e$ and $\theta_\gamma$, the
  original unbiased (non-lens) population has a flat distribution of
  $\theta_e-\theta_\gamma$. 
\label{fig:et}}
\end{figure}

As a specific example, in this paper we present the mock catalogue of
lensed QSOs and SNe expected for the baseline survey planned with
LSST (see Table~\ref{table:surveys} for the survey parameters). In
practice, the catalogue is 5 times (QSOs) or 10 times (SNe) 
over-sampled in order to reduce shot noise. The catalogue contains all
the necessary information for each mock lens, including the properties
of lenses and sources, image positions, magnifications, and time
delays between image pairs.\footnote{The mock lens catalogues
described here are available at  
\texttt{http://kipac-prod.stanford.edu/collab/research/lensing/mocklens}}

First, we examine the redshift distributions, to check the
validity of the Monte-Carlo technique. Figure~\ref{fig:zsdist}
compares the source redshift distribution of the mock lenses with the
analytic result obtained from equation (\ref{eq:dndzs}). The
reasonable agreements of the distributions assure us that the Monte-Carlo
technique is indeed feasible and reliable. The total number of strong
lenses in the mock catalogue is $3132\pm 25$ for lensed QSOs and
$122\pm4$ for lensed SNe (the errors refer to the $1\sigma$
Poisson error), which is again in good agreement with the analytic
result shown in Tables~\ref{table:lqsos} and~\ref{table:lsne}. In 
Figure~\ref{fig:zldist} we show lens redshift distributions which are
again in good agreement. 

As discussed above, the mock lens catalogue is useful in studying the
lensing bias, i.e., the difference of the lens population from the
normal population. As practical examples, in Figure~\ref{fig:et} we
show the distributions of the lens ellipticity, and of the alignment
between the lens galaxy and external shear, for double and quad
lenses separately. We find that the lensing galaxies of quad image 
systems are $\Delta e\sim 0.1$ more elliptical than those of double
lenses. This result is broadly consistent with \citet{rozo07} who
showed such a difference using the analytic approach. It appears that
there is no significant difference in the distributions of external
shear between double and quad lenses. However, we find that the
relative orientation of lensing galaxies and external shear is
biased: for quad lenses the direction of external shear is more
likely to be aligned with the major axis of the lensing galaxy,
whereas for double lenses the direction of external shear is more
likely to be aligned with the minor axis. This bias can easily be
understood by the well-known approximate degeneracy between galaxy
ellipticity and external shear \citep[see, e.g.,][]{keeton97}.

\subsection{Distributions of time delays}
\label{sec:delay}

\begin{figure*}
\begin{center}
 \includegraphics[width=0.95\hsize]{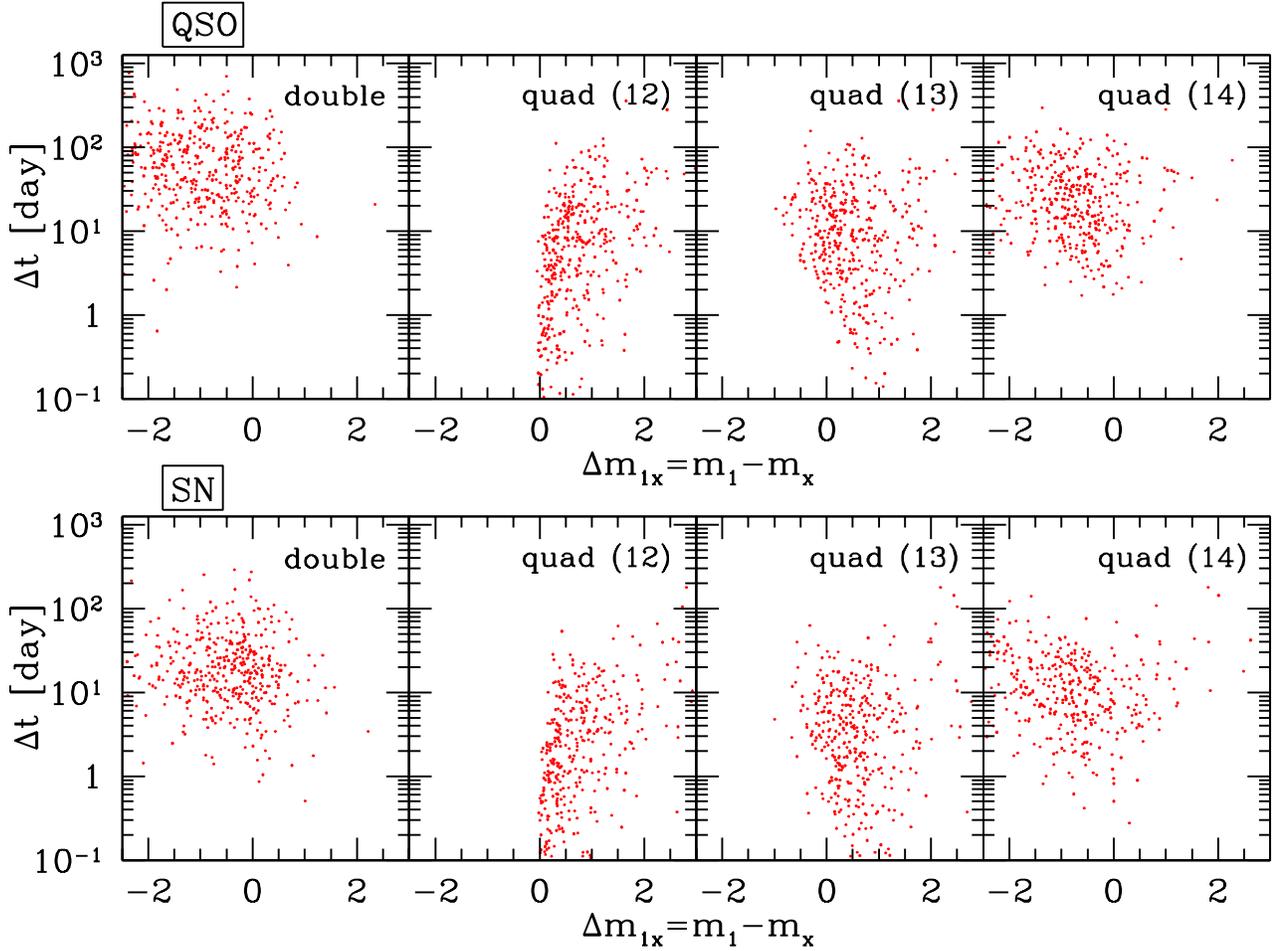}
\end{center}
\caption{Predicted distributions of time delays and magnitude
  differences for various image pairs. Each panel shows values for 400
  image pairs randomly selected from our mock lens catalogue. Here we
  name images in order of arrival, i.e., image $1$ arrives first,
  image $2$ next, etc.  The upper panels show the distributions for
  lensed quasars, whereas the lower panels are for lensed SNe.
\label{fig:dmdt}}
\end{figure*}

The predicted distributions of time delays are important in order 
to assess the measurability of time delays from future time-domain
surveys. In Figure~\ref{fig:dmdt} we show the expected distributions
of time delays and magnitude differences for strongly lensed image
pairs, derived from our LSST mock lens catalogue. For double lenses,
typical time delays are $\sim 1-3$~months. Lensed SNe have somewhat
shorter time delays on average than lensed QSOs, partly because of
their lower redshifts and the stronger effect of the magnification
bias. In most cases the brighter images arrive first, as is expected
from simple spherical mass models, but in some cases the arrival order
is inverted, i.e., the fainter images arrive first. In either case the
detection of first arrival images is assured because of our use of the
fainter image for computing the magnification bias. 

For quad lenses the situation is more complicated because there
are three independent image pairs. However, in Figure~\ref{fig:dmdt}
we can clearly see a general trend that the first arrival image tends
to be fainter than the second and third arrival images
($\Delta m_{12}>0$ and $\Delta m_{13}>0$),
and tends to be brighter than the fourth arrival image
($\Delta m_{14}<0$).
Thus we expect that the
first image to arrive is most likely the third brightest image; this
provides support for our choice of the third brightest images for
computing the magnification bias, because it assures the detection of
the first lensed SN image to arrive. The appearance of this first SN
image close to an early-type galaxy can act as an initial trigger, but
two SNe are likely to be needed to confirm the system as a lens  (and
not a pair of SNe in the lens galaxy itself). The time delay between
the first image and the subsequent images will still need to be
measured, since the delay between images 2 and 3 will often be very
short. Our result indicates that typical time delays for quad
lenses range from $\sim 1$~day for the shortest delay to $\sim
1$~month for the longest delay.

\section{Cosmological parameters from time delays in LSST} 
\label{sec:cosmo}

In this section, we use our mock LSST lens catalogue to explore one
particular application of a future time-delay lens sample: the
measurement of the Hubble constant, and also other cosmological
parameters. 

\subsection{Basic assumptions}

We follow the methodology proposed by \citet{oguri07a} to combine a
number of time delay measurements. The technique was adopted also by
\citet{coe09} to discuss future constraints from time delays.
In brief, we introduce a reduced time delay defined by
\begin{equation}
 \Xi\equiv \left|\frac{\Delta t_{ij}}{r_j^2-r_i^2}
\right|\frac{2c}{1+z_{\rm l}}\frac{D_{\rm ls}}{D_{\rm l}D_{\rm s}},
\end{equation}
where $r_i$ denotes the distance of image $i$ from the center of the
lens galaxy, and study its behavior as a function of the asymmetry
$R_{ij}=|r_j-r_i|/|r_j+r_i|$ and the opening angle
$\theta_{ij}=\cos^{-1}\left[({\mathbf x}_i\cdot{\mathbf 
 x}_i)/(r_ir_j)\right]$. While the reduced time delay $\Xi$ becomes
unity when the lens potential is exactly isothermal, external
perturbations, non-isothermality, and substructures induce scatter 
in $\Xi$. 

We construct a subsample of lensed quasars and SNe for our statistical
analysis from the mock lens catalogue. First, we restrict the image
separation range to $0\farcs8<\theta<2\farcs5$. The upper cut is meant
to remove lens systems which are significantly affected by
groups/clusters. We do not use lenses with $\theta<0\farcs8$ because
the detection and characterisation of lensing galaxies might be
difficult for such small-separation lens systems. We also limit the
lens redshift to $z_{\rm l}<1.2$ (see also Figure~\ref{fig:zldist}), beyond
which the 4000{\AA} break of lensing galaxies comes in the $z$-band
and therefore it becomes much  more difficult to study the lensing
galaxies without deep near-infrared images. Next, for simplicity we
consider only the double image lenses that dominate the LSST lens
sample. One of the reasons for this is our choice of the third
brightest images for computing magnification bias, which suggests that
we may not be able to measure some of quad time delays without deeper
follow-up monitoring. In either case, our forecast constraints are
conservative in the sense that additional quad lenses should only 
improve the constraints \citep[see also][]{coe09}. Finally, we
use only lenses with the asymmetry $0.15<R_{ij}<0.8$,
in which the effects of various perturbations on $\Xi$ are modest and
stable \citep{oguri07a}.  These cuts leave a sample of 1542 lenses
(1476 lensed QSOs and 66 lensed SNe) for our forecast study. The
sample has average lens redshift $z_{\rm l}=0.65\pm0.27$ (where we
also give the standard deviation of the sample) and average 
source redshift $z_{\rm s}=2.29\pm0.86$ (see also
Figure~\ref{fig:zldist}), which are reasonably similar to the simple
Gaussian distributions assumed in \citet{dobke09} and \citet{coe09},
although it should be kept in mind that these two redshift
distributions are not independent but are naturally correlated.

The error on $\Xi$ comes both from measurement uncertainties and from
model uncertainties \cite[see][]{oguri07a,coe09}.
We assume measurement errors of $\sigma(\Delta t)=2.0$~days
and $\sigma(r)=0\farcs01$. The assumed time delay measurement error is
the same as the one adopted by \citet{coe09}, which was based on
simulations by \citet{eigenbrod05}.  We note  that time delay
measurements in the optical waveband are complicated by microlensing
for both lensed quasars \citep[as][showed]{eigenbrod05} and lensed
supernovae. \citet{dobler06} discuss the extent to which  microlensing
can bias time delay measurements in lensed supernovae. While we should
be optimistic about achieving 2-day precision given the wealth of
additional imaging and  spectroscopic data we can anticipate gathering
on these precious systems, this is a topic requiring further research.

Based on the Monte-Carlo simulations of
\citep{oguri07a}, we adopt the model uncertainty of
$\sigma(\log\Xi)=0.08$ which comes mostly from external shear and the
scatter in the radial density slope. In this paper, we assume no error
on the lens and source redshifts, which corresponds to the situation
where all redshifts are measured spectroscopically. See \citet{coe09}
for the effect of photometric redshift errors on cosmological parameter
estimates, which appears to be modest compared with lens potential
uncertainties. 

As discussed in \citet{oguri07a}, the main source of systematic error
is the unknown mean logarithmic slope of the radial density profile of
the lensing galaxies. The
 parameter~$\alpha = \partial\log{\phi}/\partial\log{r}$ for each lens
is the effective slope of the lens potential, that
includes the effect of the external convergence ($\alpha=1$ 
corresponds to an effectively isothermal potential, 
$\phi({\mathbf x})\propto r^\alpha$). We include the mean slope
$\bar{\alpha}$ as an additional nuisance parameter. We include the
effect of changing $\bar{\alpha}$ by adopting the scaling relation, 
$\Xi \propto \Delta t_{ij} \propto (2-\bar{\alpha})$ 
\citep[e.g.,][]{refsdal94,witt00,wucknitz02},  
and assign it a Gaussian prior. We adopt the fiducial value of 
$\sigma_{\rm prior}(\bar{\alpha})=0.005$, and marginalise over 
$\bar{\alpha}$ to obtain constraints on cosmological parameters. 
We explicitly explore the impact of this assumption in 
Section~\ref{sec:cosmo:results}.

\subsection{Fisher matrix analysis}

We obtain expected constraints on cosmological parameters using the
Fisher matrix approach. We compute chi-squared as
\begin{equation}
\chi^2=\sum_{\rm lens}\frac{[\log \Xi(z_{\rm l}, z_{\rm s})-\log \Xi_{\rm
      mock}]^2}{\sigma(\log \Xi)^2}.
\end{equation}
Here $\sigma(\log \Xi)$ is the quadratic sum of measurement
and model uncertainties. The summation is over the 1542 mock
lenses. Then the Fisher matrix for the parameter set \{$p_1$, $p_2$,
  $\ldots$\} is calculated as
\begin{equation}
F_{\mu\nu}^{\rm lens}=\frac{1}{2}\frac{\partial^2 \chi^2}{\partial p_\mu
  \partial p_\nu}.
\end{equation}

Here we adopt the standard parametrisation of the dark energy equation of
state of the form: 
\begin{equation}
w(a)=w_0+(1-a)w_{\rm a}.
\end{equation}
To complement the strong lensing data, we include the expected
constraints from Cosmic Microwave Background (CMB) measurements and
distance-redshift relation measurements from (unlensed) type-Ia
supernovae. We compute the CMB Fisher matrix $F_{\mu\nu}^{\rm CMB}$
from upcoming Planck data using the simplified approach described in
\citet{mukherjee08}.  In the LSST era, we can expect this to be the
basic prior PDF for dark energy studies \citep{albrecht06}.
For the SNe, we estimate expected constraints from a JDEM/SNAP survey
assuming the redshift distribution of SNe in the final subsample given
in \citet{aldering04}. 
Following \citet{hu06}, we add 300 local supernovae at $z=0.05$, and
include both statistical and systematic errors in computing the Fisher
matrix $F_{\mu\nu}^{\rm SN}$. In doing so, we include a nuisance
parameter, $\mathcal{M}$, which corresponds to the absolute magnitude
of type-Ia supernovae, and marginalise over it. 

To summarise, assuming a flat universe, we consider the following 8
parameters: \{$w_0$, $w_{\rm a}$, $\Omega_{\rm m}$, $h$, $\bar{\alpha}$, 
$\Omega_bh^2$, $n_s$,
$\mathcal{M}$\}. 
Note that the  time delays depend only on the first 5 of these
parameters. Once the Fisher matrices are calculated, we obtain the
parameter covariance matrix as 
$C_{\mu\nu}=({\mathbf F}^{-1})_{\mu\nu}$, and display the 
corresponding marginalised constraints as 
ellipses containing 68\% of the probability implied. 

\begin{figure}
\begin{center}
 \includegraphics[width=0.95\hsize]{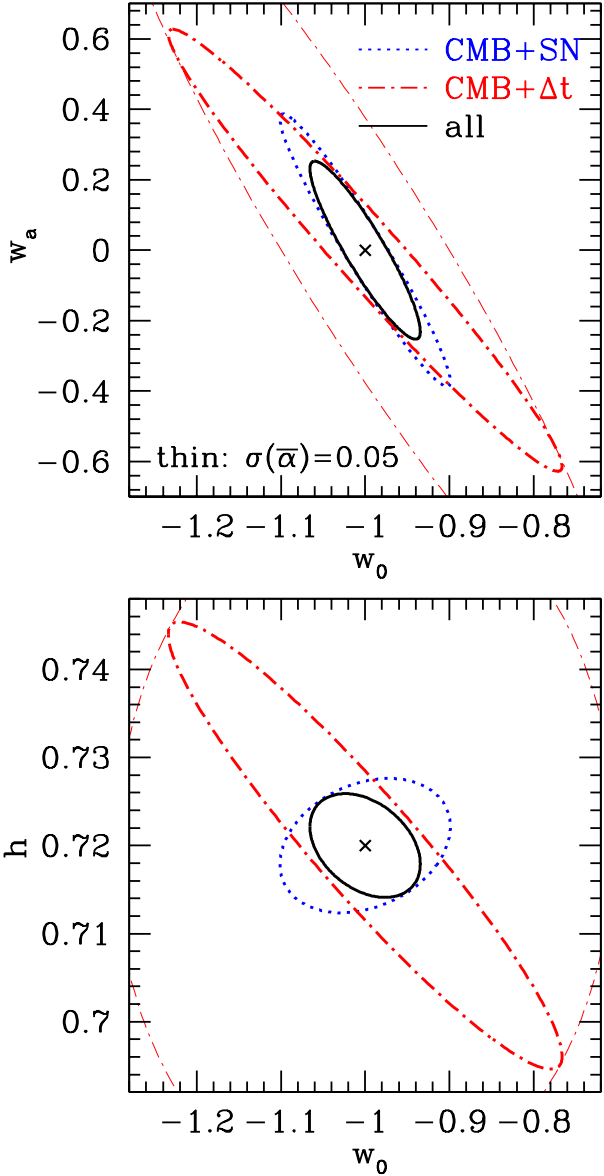}
\end{center}
\caption{Forecast marginalised constraints in the $w_0$-$w_{\rm a}$
  plane ({\it upper panel}) and the $w_0$-$h$ plane ({\it lower panel}). 
  Ellipses show 68\% confidence regions for various combinations of expected
  constraints from the Planck satellite (CMB), type-Ia supernovae in
  SNAP (SN), and time delays in LSST ($\Delta t$). Thin ellipses are
  constraints from CMB+$\Delta t$, but assumes a much weaker mean
  slope prior of $\sigma_{\rm prior}(\bar{\alpha})=0.05$ instead of
  the fiducial value of $0.005$.
\label{fig:cons}}
\end{figure}

\subsection{Results}
\label{sec:cosmo:results}

We show the marginalised constraints on the cosmological parameters of
interest in Figure~\ref{fig:cons}. 
We find that the LSST strong lens time delays, 
combined with the Planck constraints,
provide marginalised $1\sigma$ uncertainties (68\% confidence
intervals) of $\sigma(w_0)=0.15$, 
$\sigma(w_{\rm a})=0.41$, and  $\sigma(h)=0.017$. 
If we include the JDEM/SNAP SNe constraints, 
we obtain $\sigma(w_0)=0.04$,
$\sigma(w_{\rm a})=0.17$, and $\sigma(h)=0.004$. Although the
constraints from
time delays are less tight than those from the SNe, 
time delays still
improve the constraint on the dark energy parameters
because of the different degeneracy
directions. This point is particularly clear in the $w_0$-$h$ plane,
where the two constraints are almost perpendicular with each other. 
One way to view this is that future supernova surveys
can be enhanced by strong lens time delay distance measurements.

We also consider the case where we have a 
much weaker prior on the mean
radial density slope $\bar{\alpha}$: if the prior PDF for this
parameter is ten times broader
than in our fiducial case, $\sigma_{\rm prior}(\bar{\alpha})=0.05$, 
the resulting constraints from time delays become much weaker
(see Figure~\ref{fig:cons}) and hardly improve future constraints on
dark energy from type-Ia SNe. This highlights the importance of
accurate prior knowledge on lens potentials when deriving cosmological
constraints from time delays. The slight rotation of the ellipse when
the prior on $\bar{\alpha}$ is broadened is a result of the time delay
dataset contributing less strongly to the joint fit -- lack of
knowledge of the mean density profile slope leads to lensing being
downweighted as a cosmological probe. 

\begin{figure}
\begin{center}
 \includegraphics[width=0.95\hsize]{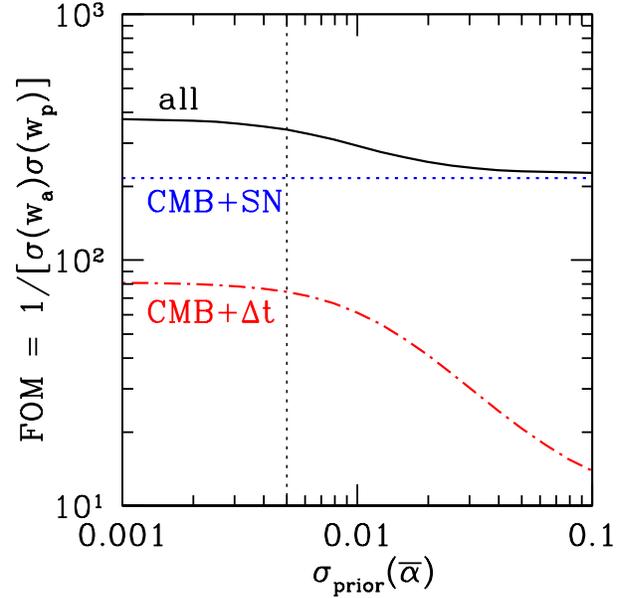}
\end{center}
\caption{Dark energy figure of merit \citep[FOM; e.g.,][]{albrecht06},
  defined as $1/\left[\sigma(w_{\rm a})\sigma(w_p)\right]$ with $w_p$ being
  equation of state at the pivot redshift, as a function of the input
  (prior) uncertainty on the mean radial lens density profile
  slope $\bar{\alpha}$. The vertical dotted lines indicate our
  fiducial prior value, $\sigma_{\rm prior}(\bar{\alpha})=0.005$. 
\label{fig:fom}}
\end{figure}

Indeed, the results presented here depend crucially on the
width 
of the prior on the mean slope, $\sigma_{\rm prior}(\bar{\alpha})$.
In Figure~\ref{fig:fom}, we show how the constraints on dark 
energy are
affected by changing $\sigma_{\rm prior}(\bar{\alpha})$. We adopt the
standard method of quantifying the constraints on the dark energy
parameters, employing the Dark Energy Task Force ``figure of merit''
(DE FOM), which is proportional to the inverse of the ellipse area in
the $w_0$-$w_{\rm a}$ plane \citep{albrecht06}. 
We find that  time delays would
not improve upon the constraints from a JDEM/SNAP SNe survey if
$\sigma_{\rm prior}(\bar{\alpha})\gg0.01$, which suggests that we need
to determine the mean radial slope of galaxy density profile to better
than $\sim 0.01$ in order to use time delays in LSST as a future
cosmological probe. In the limit of perfect prior information
on~$\bar{\alpha}$, we see that the LSST time delay sample could
improve the DE FOM by a factor of 1.74, from 215 to 375.

\section{Discussion} 
\label{sec:discuss}

\subsection{Static surveys}

In this paper, we focused on cadenced surveys, mainly because of the
interest in strong lens time delays for use in measuring cosmological
parameters, and the  promising technique to find such  lenses from
their time variability \citep{kochanek06b}. As discussed in
\S\ref{sec:surveys}, however, some future wide-field optical imaging
surveys will lack sufficient  time domain sampling frequency to
measure time delays. These surveys will still be useful for locating
many lensed quasars, and possibly lensed SNe as well, but they will
require extensive follow-up monitoring observations to measure the
time delays.

We note that some planned optical imaging surveys will be completely
static, i.e., observing each field only once. Previous lens surveys,
however, have convincingly shown that we can still identify lensed
quasars by checking the colour and morphology of quasars
\citep[e.g.,][]{inada08}. High resolution, space-based surveys will be 
particularly effective here. For instance, the planned space mission 
Euclid \citep{refregier10} is planning to obtain high-resolution
imaging data for the entire southern extragalactic sky, 
which will be enormously
useful for locating many lensed quasars from morphology. Once these quasar
lens candidates have been identified, follow-up images can be taken to assess
the variability of the candidates and to improve the efficiency of
lens finding. Again, we will need additional monitoring data in order to
measure time delays. 
Monitoring follow-up of lenses found
in static surveys will, however, represent a significant observational 
undertaking.

\subsection{Follow-up observations}

Even for well-cadenced surveys such as LSST, we can conduct follow-up
monitoring to obtain more accurate time delays. In some cases the
survey duration will not be enough to measure long time delays, and
for these systems additional monitoring after the survey will be
necessary.  Particular attention should be paid to the follow-up of
strongly lensed SNe because of their transient nature. Clearly it is
important to identify lensed SNe as early as possible. In principle it
will be possible to trigger follow-up monitoring by the appearance of
the first lensed SN image, by checking the difference of photometric
redshifts between the SN and its nearby galaxies, or from prior
knowledge of the lensed host galaxies. Note that good image position
and colour information for SN classification and redshift estimation
will be required for this to work. Where these are not available,
confirmation of a lensed SN will likely require the appearance of the
second image in a lensing configuration. 

In our Fisher matrix analysis, we assumed that all the redshifts are
known. Spectroscopic redshifts for lensed quasars are relatively easy
to obtain, thanks to prominent quasar emission lines. However,
we can expect spectroscopy of the lensed SNe and the lensing galaxies
to be generally much more difficult, because one has to detect
absorption lines to measure the redshifts. Instead, we can adopt
photometric redshifts, provided we can cleanly separate the lensed
images from the lens galaxy light; if there is no bias inherited in
the photometric redshift measurement, its effect is just to degrade
the cosmological constraints from the time delays slightly 
\citep[see][]{coe09}. Spectroscopic follow-up observations are also
helpful for the confirmation of strong lens systems, although the
detection of time delays between the images does itself serve as a
strong test of the lensing hypothesis.

\subsection{Prospect for constraining lens potentials}

One of our main findings (\S\ref{sec:surveys}) is that accurate
prior knowledge on lens potentials is crucial for cosmological
constraints from time delays. Among others, the most important
parameter is the mean slope of the effective radial density profile of
the lensing galaxy, $\bar{\alpha}$, including the effect of external
convergence. Our requirement of $\bar{\alpha}\la 0.01$ for the LSST
time delay sample to be a useful addition to the cosmographic toolbox
is in fact not far from the current statistical error of $\sim 0.02$
from the Sloan Lens ACS Survey \citep{koopmans09}. Care will need to
be taken to ensure that the LSST lens galaxy sample is similarly well
characterised. In addition, we expect that the external convergence
may be well modelled by combining photometric information for the
surrounding field with ray-tracing in $N$-body simulations
\citep{oguri05,suyu10}. 

With such a large number of strong lens systems, it may be possible to 
determine cosmological parameters and the mean radial density slope
simultaneously. For instance, \citet{rusin05} constrained the mean
radial density profile by combining the Einstein radius measurements
for lenses with different source redshifts. Moreover, we may also be
able to use flux ratios between images to constrain the radial profile
\citep[e.g.,][]{mortsell06,mutka10}, although we will 
need to take into account the effects of dark halo substructures,
differential dust extinction, and microlensing very carefully. We
leave the exploration of these possibilities to future work. 

\subsection{Comparison with individual lens modelling}

While we considered a statistical analysis for combining
many time delay measurements, tight cosmological constraints may also
be obtained from the detailed modelling of individual lens systems (the 
so-called ``golden lens'' approach). Indeed, the feasibility of this
has been nicely demonstrated, by e.g. \citet{suyu10} for B1608$+$656
and \citet{fadely10} for Q0957$+$561. One advantage of our statistical
approach is that it can handle various perturbations on lens
potentials, such as substructures and higher-order external
perturbations, more straightforwardly, since their effects average out
when combining many lenses. 
Although the average does not always converge to zero (an example is
external convergence), this poses no problems as long as the residual
can be predicted reasonably accurately by theory.
Indeed, the statistical framework provides a means, in principle, to
infer the statistical properties of lens substructure, microlensing
and dust simultaneously with the cosmology and lens profile
parameters. 

For individual modelling we do not need to know the mean radial
density profile very accurately, but instead need enough observational 
constraints to determine the radial profiles of each individual 
lensing galaxy. For this reason, we expect the quad lenses to be the
focus of the individual modelling efforts, complenting the preference
for doubles in the statistical approach. To summarise, we expect that
individual mass modelling will provide important complementary
constraints on cosmological parameters, and help to assess the 
systematic errors that each methods involves.

\subsection{Group- or cluster-scale lenses}

Future surveys will discover not only galaxy-scale lenses but also
wide-separation lenses produced by massive groups or clusters of
galaxies, such as those discovered in the SDSS \citep{inada03,inada06}.
The time variability information will also be helpful in identifying
these lenses as well, but the time delays between images will 
tend to be long, $\sim 1-10$~years for massive clusters, suggesting
that well-designed follow-up monitoring program may be necessary to
measure their time delays.

\section{Conclusions} 
\label{sec:concl}

We have presented detailed calculations of the likely yields of
several planned cadenced surveys, adopting realistic distributions for
the lens and source properties, and taking account of the selection
functions, including the magnification bias. We find that, for
example, the LSST will discover $\sim 8000$ lensed quasars ($\sim
3000$ of which will have time delay measurements) and $\sim 100$
lensed SNe. Approximately one third of lensed SNe will be type-Ia. The
lenses are dominated by double (two-image) lenses, with expected quad
fractions of $\sim 14$\%  for lensed quasars and $\sim 30$\% for lensed
SNe. We have also produced mock catalogues of lenses, which are 
useful for probing strong lensing selection effects and the
feasibility of various science projects.  

We have used a mock catalogue of $\sim 1500$ well-observed double
lenses in LSST to derive expected cosmological constraints. 
Specifically, we derived precisions on the Hubble constant and the
dark energy equation of state 
parameters from this sample of time delay
measurements, assuming priors from Planck. The resulting predicted
marginalised 68\% confidence intervals are $\sigma(w_0)=0.15$,
$\sigma(w_{\rm a})=0.41$, and $\sigma(h)=0.017$, implying that 
time delays can improve constraints from a JDEM supernova type-Ia
sample. However, this result holds only if we have
accurate prior knowledge of the 
lens population's mean effective 
density profile: we find that the prior on the mean radial density
slope has to be $\bar{\alpha}\la0.01$ in order for time delays from
the LSST strong lens to be a useful future cosmological probe.

\section*{Acknowledgments}

We thank 
C.~Fassnacht, C.~Keeton, G.~Dobler, R.~Blandford, H.~Zhan, T.~Tyson, 
S.~Jha, P.~Nugent, 
M.~Strauss, Z.~Ivezi\'{c}, S.~Smartt, D.~Young, E.~Buckley-Geer, 
J.~Annis, 
T.~Hamana, N.~Yasuda, and M.~Sullivan, 
for useful discussions.
We also thank an anonymous referee for helpful suggestions.
This work was supported by Department of Energy contract
DE-AC02-76SF00515.
PJM acknowledges support from the TABASGO and Kavli foundations, 
in the form of research fellowships.
 


\label{lastpage}

\end{document}